\documentclass[useAMS,usenatbib]{mn2e}
\usepackage{rotating}
\usepackage{journals}

\def\approxgt{\ifmmode \rlap{$>$}{}_{{}_{{}_{\textstyle\sim}}} \else%
$\rlap{$>$}{}_{{}_{{}_{\textstyle\sim}}}$\fi} 
\def\approxlt{\ifmmode \rlap{$<$}{}_{{}_{{}_{\textstyle\sim}}} \else%
$\rlap{$<$}{}_{{}_{{}_{\textstyle\sim}}}$\fi}

\def\farcs{\hbox{$.\!\!^{\prime\prime}$}}

\def\arcmin{\hbox{$^\prime$}}
\def\arcsec{\hbox{$^{\prime\prime}$}}

\LARGE \normalsize \title[1M~1716--315 in quiescence]{The quasi--persistent neutron star soft X--ray
transient 1M~1716--315 in quiescence}

\author[Jonker et al.]  {P.G.~Jonker$^{1,2,3}$\thanks{email :
p.jonker@sron.nl}, C.G. Bassa$^3$, S.~Wachter$^{4}$\\
$^1$SRON, Netherlands Institute for Space Research, Sorbonnelaan 2, 3584~CA, Utrecht, The Netherlands\\
$^2$Harvard--Smithsonian  Center for Astrophysics, 60 Garden Street, Cambridge, MA~02138, Massachusetts,
U.S.A.\\
$^3$Astronomical Institute, Utrecht University, P.O.Box 80000, 3508 TA, Utrecht, The Netherlands\\
$^4$Spitzer Science Center, California Institute of Technology, 1200
E.~California Blvd., Pasadena CA 91125, U.S.A. \\
}

\begin{document}

\maketitle

\begin{abstract} \noindent We report on our analysis of a 20 ksec {\it Chandra} X--ray observation
of the quasi--persistent neutron star soft X--ray transient (SXT) 1M~1716--315 in quiescence. Only
one source was detected in the HEAO--I error region. Its luminosity is
1.6$\times10^{32}-1.3\times10^{33}$ erg s$^{-1}$. In this the range is dominated by the uncertainty
in the source distance. The source spectrum is well described by an absorbed soft spectrum, e.g.~a
neutron star atmosphere or black body model.  No optical or near--infrared counterpart is present
at the location of the X--ray source, down to a magnitude limit of $I\approxgt 23.5$ and
$K_s\approxgt 19.5$. The positional evidence, the soft X--ray spectrum together with the optical
and near--infrared non--detections provide strong evidence that this source is the quiescent neutron
star SXT. The source is 10--100 times too bright in X--rays in order to be explained by stellar
coronal X--ray emission. Together with the interstellar extinction measured in outburst and
estimates for the source distance, the reported optical and near--infrared limit give an upper
limit on the absolute magnitude of the counterpart of $I>8.6$ and $K_s>5.1$. This implies that the
system is either an ultra--compact X--ray binary having ${\rm P_{orb}<1\,hr}$ or the companion star
is an M--dwarf. We reconstructed the long term X--ray lightcurve of the source. 1M~1716--315 has
been active for more than 12 years before returning to quiescence, the reported {\it Chandra}
observation started 16.9$\pm$4.1 years after the outburst ended.

\end{abstract}

\begin{keywords} stars: individual (1M~1716--315) --- 
accretion: accretion discs --- stars: binaries --- stars: neutron
--- X-rays: binaries
\end{keywords}

\section{Introduction} 

Low--mass X--ray binaries (LMXBs) are highly evolved, interacting binaries in which a neutron star or
black hole accretes matter via an accretion disc which is fed by a cool, low--mass star (typically
M$\approxlt 1$ M$_\odot$). Some systems have an orbital period of less than $\sim$1 hour (ultra--compact
X--ray binaries; UCXBs). These ultra--compact systems are so compact that the donors cannot be main
sequence stars, but instead must be hydrogen--poor (semi) degenerate stars (e.g.~\citealt{verbunt1995}).
\citet{2004ApJ...603..690B} predict that about half of the total population of LMXBs may be in UCXBs. A
large fraction of the LMXBs are found to be transient systems -- the so called soft X--ray transients
(SXTs; e.g.~see \citealt{1997ApJ...491..312C}).

A special group of soft X--ray transients has several--year long outbursts before they return
to quiescence. These quasi--persistent systems provide an ideal testing ground since the
accretion history is well--constraint. Furthermore, the idea is that the long outburst has
heated the neutron star crust to temperatures higher than that of the neutron star core.
Therefore, the (evolution of the) quiescent X--ray luminosity provides information on the
cooling properties of the neutron star, which in turn depend on the neutron star equation of
state (\citealt{1998ApJ...504L..95B}; \citealt{2001ApJ...548L.175C};
\citealt{2001ApJ...560L.159W}; \citealt{2002ApJ...580..413R}; \citealt{2006MNRAS.372..479C}).

1M~1716--315 was first detected in OSO--7 observations obtained between Sept.~29, 1971 and May 18,
1974 (\citealt{1975IAUC.2765Q...1M}). The last reported detection was by
\citet{1988MNRAS.232..551W} using EXOSAT observations obtained in June 14--18, 1984. The source was
not detected by the ROSAT all sky survey (\citealt{1999A&A...349..389V}). Large X--ray flares
lasting $\approx$10 minutes have been reported (\citealt{1976ApJ...208L.115M}). These and other
(shorter) flares reported by \citet{1978Natur.272..701J} would now most likely be classified as
type~I X--ray bursts. The confirmation of the occurence of type~I X--ray bursts and hence of the
neutron star nature of the compact object in 1M~1716--315 was given by the detection of three
bursts by the Hakucho satellite (\citealt{1981ApJ...244L..79M}). Later Hakucho detected a long
radius expansion burst (\citealt{1984ApJ...276L..41T}). The properties of this burst were used by
\citet{1984PASJ...36..861T} to derive a distance of $\sim 6$ kpc (\citealt{2004MNRAS.354..355J}
derived d=5.1--6.9 kpc using the properties of the same burst; the range depends on whether the
burst was hydrogen or helium rich).

The source position was determined to 90\arcsec~by SAS--3 (90 per cent confidence;
\citealt{1978Natur.272..701J}). The position was later refined by HEAO--I to an area with an
equivalent circular radius of 23\arcsec~(90 per cent confidence;
\citealt{1980AJ.....85.1062R}). The extinction towards the source is low for an LMXB;
\citet{1997ApJS..109..177C} report an equivalent hydrogen column density of
N$_H=(2.1\pm0.6)\times 10^{21}$ cm$^{-2}$. \citet{2005AIPC..797..639W} observed the source for
1 ksec.~with the High Resolution Camera onboard the {\it Chandra} satellite but did not detect
the source; they report an upper limit on the flux of 2.8$\times10^{-14}$ erg cm$^{-2}$
s$^{-1}$. In this Manuscript, we report on a deeper {\it Chandra} observation of the neutron
star soft X--ray transient 1M~1716--315 in quiescence. We further report on optical Magellan
and near--infrared Blanco observations obtained while the source was in quiescence.

\section{Observations, analysis and results} 

\subsection{X--ray observations}

\begin{figure}
  \includegraphics[angle=0,width=8cm,clip]{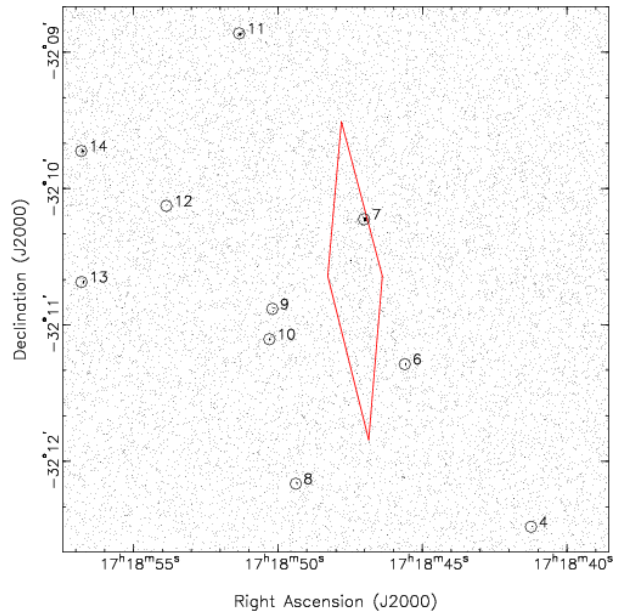}
  
  \caption{ Part (4\arcmin$\times$4\arcmin) of {\it Chandra}'s ACIS--S3 CCD is
  shown. The diamond shaped region is the HEAO--1 error region of 1M~1716--315.
  Only one of the detected sources is consistent with that position. The circles
  indicate the positions of detected X--ray sources (see
  Table~\ref{tab:srces} for the numbering).}

\label{fig:xsrc}
\end{figure}

We observed 1M~1716--315 with the back--illuminated S3 CCD--chip of the Advanced
CCD Imaging Spectrometer (ACIS) detector on board the {\it Chandra} satellite.
The observations started on MJD~53567.604 (UTC; July 16, 2005). The net,
on--source exposure time was $\sim$20 ks. The data telemetry mode was set to
\textsc {very faint} to allow for a better background subtraction. After the
data were processed by the {\it Chandra} X--ray Center (ASCDS version 7.6.0), we
analysed them using the \textsc {ciao 3.3} software developed by the {\it Chandra}
X--ray Center. We reprocessed the data to clean the background taking full
advantage of the \textsc {very faint} data mode. We searched the data for
background flares but none were found, hence we used all data in our analysis.
Using the \textsc {ciao} tool  \textsc {wavdetect} we detect and determine
positions of 20 sources in the field of view of the ACIS S3 CCD (see
Fig.~\ref{fig:xsrc} and Table~\ref{tab:srces}).

\subsection{Optical observations}

In order to search for the optical counterpart in quiescence, we have obtained
$I$--band images with the Inamori--Magellan Areal Camera and Spectrograph
(IMACS) instrument mounted on the 6.5~m Magellan--Baade telescope on July 7,
2005, 00:00:35 UTC (MJD 53558.0004 UTC). Exposure times of 10~seconds and 2x300
seconds were used. IMACS is a mosaic of eight 4k$\times$2k CCDs that were
operated in a 2$\times$2 binning mode. The seeing was 0\farcs7. We calibrated
all 8 CCD chips of the 10\,s $I$-band image against stars from the second
version of the USNO CCD Astrograph Catalog (UCAC2;
\citealt{2004AJ....127.3043Z}). Between 13 and 28 stars were used to compute the
transformation for each chip, giving typical root--mean--square residuals of
$0\farcs058$ and $0\farcs051$ on $\alpha$ and $\delta$, respectively. From the
20 X--ray sources found on the ACIS--S3 chip, 8 coincided with relatively bright
stars on the 10\,s $I$--band image (see Table~\ref{tab:srces}) having $I$--band
magnitudes between 11 and 16. The celestial positions of these stars were
measured from the calibrated images and compared with the X--ray positions to
determine a boresight correction of $\Delta \alpha=-0\farcs001\pm0\farcs037$ and
$\Delta \delta=+0\farcs016\pm0\farcs035$. Using this boresight correction we
obtained positions and uncertainties for the 20 sources detected by {\it
Chandra} on the S3 CCD chip (see Table~\ref{tab:srces}).  

Subsequently, the astrometric solution of the 10--second frame was transferred
to the 300--second images using $\sim$1700 stars. In this the rms uncertainty
was 0\farcs037 in each coordinate. For the X--ray source in the HEAO--I error
region the uncertainty in its location due to the wavelet fitting in \textsc
{wavdetect} is 0\farcs045 in $\alpha$ and 0\farcs038 in $\delta$. Adding all the
errors in quadrature gives at the 99\% confidence level an radius for the error
circle of $0\farcs36$. The images have been corrected for bias and flatfielded
with \textsc {iraf}\footnote{\textsc {iraf} is distributed by the National
Optical Astronomy Observatories}.

\begin{table*} \caption{Names, coordinates and detection significance of the 20 sources
detected by {\it Chandra} on the S3 CCD chip sorted on Right Ascention.}

\label{tab:srces}
\begin{tabular}{llll}
\hline
Number \& Name & $\alpha$ (h:m:s$\pm$\arcsec, J2000.0)& $\delta$ ($^\circ$:\arcmin:\arcsec$\pm$\arcsec, J2000.0) & $\sigma$\\ 
\hline
1  CXOU~J171830.7-320927$^c$ & 17:18:30.69$\pm$0.18  &-32:09:27.31$\pm$0.13  &11.9\\
2  CXOU~J171833.1-320630$^{b,c}$ & 17:18:33.10$\pm$0.10  &-32:06:30.85$\pm$0.08  &44.2\\
3  CXOU~J171836.0-320940$^c$ & 17:18:36.01$\pm$0.13  &-32:09:40.40$\pm$0.11  &17.1\\
4  CXOU~J171841.2-321228 & 17:18:41.24$\pm$0.19  &-32:12:28.90$\pm$0.19  &5.1\\
5  CXOU~J171842.4-320643$^c$ & 17:18:42.40$\pm$0.32  &-32:06:43.74$\pm$0.18  &7.1\\
6  CXOU~J171845.6-321117 & 17:18:45.61$\pm$0.12  &-32:11:17.32$\pm$0.17  &7.1\\
7  CXOU~J171847.0-321013$^a$ & 17:18:47.02$\pm$0.08  &-32:10:13.54$\pm$0.07  &41.9\\
8  CXOU~J171849.4-321209 & 17:18:49.38$\pm$0.14  &-32:12:09.93$\pm$0.13  &7.8\\
9  CXOU~J171850.2-321053 & 17:18:50.19$\pm$0.16  &-32:10:53.03$\pm$0.19  &4.8\\
10  CXOU~J171850.3-321106 & 17:18:50.30$\pm$0.20  &-32:11:06.35$\pm$0.22  &4.8\\
11  CXOU~J171851.3-320851$^b$ & 17:18:51.34$\pm$0.15  &-32:08:51.77$\pm$0.11  &12.3\\
12  CXOU~J171853.9-321007 & 17:18:53.86$\pm$0.23  &-32:10:07.64$\pm$0.24  &4.6\\
13  CXOU~J171856.8-321041$^b$ & 17:18:56.80$\pm$0.14  &-32:10:41.14$\pm$0.14  &9.5\\
14  CXOU~J171856.8-320943$^b$ & 17:18:56.81$\pm$0.13  &-32:09:43.55$\pm$0.14  &11.2\\
15  CXOU~J171858.0-320932$^c$ & 17:18:58.00$\pm$0.27  &-32:09:32.44$\pm$0.28  &4.6\\
16  CXOU~J171859.2-321147$^{b,c}$ & 17:18:59.22$\pm$0.10  &-32:11:47.86$\pm$0.10  &16.1\\
17  CXOU~J171900.2-320958$^{c}$ & 17:19:00.25$\pm$0.23  &-32:09:58.98$\pm$0.22  &5.2\\
18  CXOU~J171903.3-321123$^{b,c}$ & 17:19:03.30$\pm$0.11  &-32:11:23.40$\pm$0.11  &26.8\\
19  CXOU~J171903.4-320925$^{b,c}$ & 17:19:03.35$\pm$0.20  &-32:09:25.24$\pm$0.21  &7.6\\
20  CXOU~J171905.0-321023$^{b,c}$ & 17:19:04.98$\pm$0.11  &-32:10:23.14$\pm$0.11  &21.5\\
\end{tabular}

{\footnotesize$^a$ 1M~1716--315 in quiescence.}\\
{\footnotesize$^b$ Bright optical star at the X--ray position used for boresight correction. }\\
{\footnotesize$^c$ This source is outside the field shown in Figure 1.}\\

\end{table*}

\begin{figure}
  \includegraphics[angle=0,width=8cm,clip]{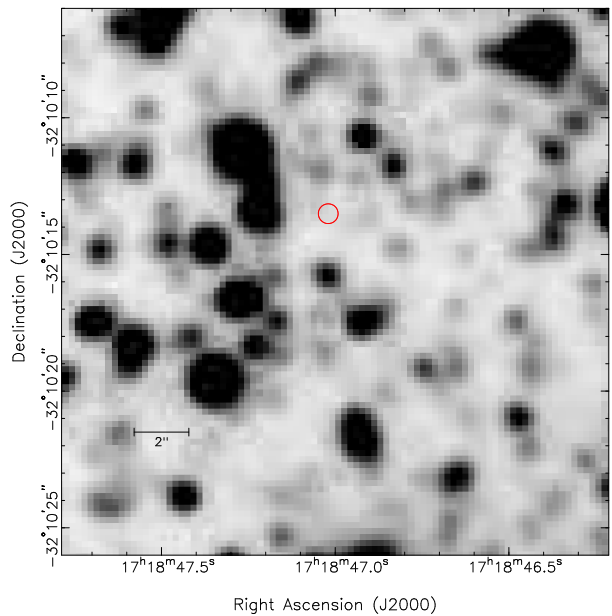}
  
  \caption{An $I$--band finder chart of the field of 1M~1716--315 obtained with Magellan/IMACS (North
  is up and East is left). The small circle indicates the position of the {\it Chandra} detected
  X--ray source number 7 (radius error circle 0\farcs36, 99 per cent confidence).}

\label{fig:src}
\end{figure}

\subsection{Near--infrared observations} 

We have also obtained near--IR $K_s$ band data with the Blanco 4~m telescope and Infrared Side
Port Imager (ISPI) at Cerro Tololo Inter--American Observatory (CTIO) on  2004 July 25 UTC. The
observation consists of three 20 sec coadds  at nine dither positions.  The data were
flatfielded, sky background subtracted and  combined into a single final image in the standard
manner using \textsc{iraf}.  ISPI has a pixel scale of  0\farcs3\ pixel$^{-1}$ and the seeing
conditions during the observations varied between 0\farcs9--1\farcs0. We have derived a
photometric calibration transformation for the final combined image based on a comparison to
2MASS measurements, neglecting any colour terms since we only have data in one filter. The
formal rms error in the transformation fit was 0.04 mag.

\subsection{The quiescent X--ray counterpart to 1M~1716--315?}

Only one source is detected in the HEAO--I error region of 1M~1716--315 (source number 7 in
Table~\ref{tab:srces}). Hence, source 7 is a strong counterpart candidate. In the error circle of source
7 on the optical $I$--band image there is no object present. We estimate a $5\sigma$ detection limit of
$I>23.5$ (see Jonker et al.~2006). Furthermore, no $K_s$ counterpart is detected in the Chandra error
circle and we estimate a detection limit of $K_s>$19.5. There are three other sources detected in X--rays
that have a position close to the error region (sources 6, 9 and 10 using the numbering from
Table~\ref{tab:srces} in Figure~\ref{fig:xsrc}). For none of these sources we detected the $I$-- or
$K_s$--band counterpart. In the case of the X--ray source number 9 a bright source was present both in
the $I$-- and the $K_s$--band image at a location close to, but not consistent with, its {\it Chandra}
error circle. The wings of the psf of the bright star, however, cover the {\it Chandra} error circle of
source number 9 raising the level of the background considerably. This precludes us from putting firm
limits on the presence of a star at the position of the X--ray source number 9. Hence, on the basis of
the optical to X--ray luminosity ratio one can not rule out any of these sources as the potential
counterpart of 1M~1716--315. Furthermore, the X--ray source in the error circle could in principle be
unrelated to 1M~1716--315: the quiescent X--ray counterpart of 1M~1716--315 could be fainter than our
detection limit. In a 1\arcsec~circle, placed randomly in the HEAO--I error region we find 0--3 counts.
Using the conservative limit of 3 counts and \citep{1986ApJ...303..336G} this converts to a 95 per cent
count rate upper limit of 3.9$\times10^{-4}$ counts s$^{-1}$. For a power law with index 2 and an
N$_H=2.1\times10^{21}$ cm$^{-2}$ this yields a 3$\times10^{-15}$ erg cm$^{-2}$ s$^{-1}$ limit on the
0.5--10 keV flux. For a distance of 6.9 kpc this gives a limit on the 0.5--10 keV luminosity of
2$\times10^{31}$ erg s$^{-1}$.

\subsection{The X--ray spectrum and lightcurve}

We extracted the X--ray photons from a circular region with a diametre of 3\arcsec~of source 7, 6, 9
and 10. For source 7, background photons were extracted from an annulus centred on the source with an
inner and outer radius of 10\arcsec~and 40\arcsec, respectively. For source 7 we detect 185 source
photons in the 19,924--seconds long effective exposure. Hence, the source count rate is
$(9.3\pm0.7)\times 10^{-3}$ counts s$^{-1}$. The lightcurve of the source is consistent with being
constant.  At these low background--subtracted count rates, photon pile--up is unimportant. We searched
for spectral variability during the observation by comparing the mean energy and the variance therein
of the photons in the first half of the observation with that of the second half. The two values
are consistent with being the same. Even though the shape of the spectrum can have changed keeping the
mean energy and its variance constant, it is more likely that within the accuracy of our data the
spectrum did not change during the observation. In order to validate the use of the $\chi^2$ fitting
technique in the spectral analysis, the source spectrum was rebinned such that each of the spectral
bins contained at least 10 counts. 

\begin{figure}
  \includegraphics[angle=-90,width=7cm,clip]{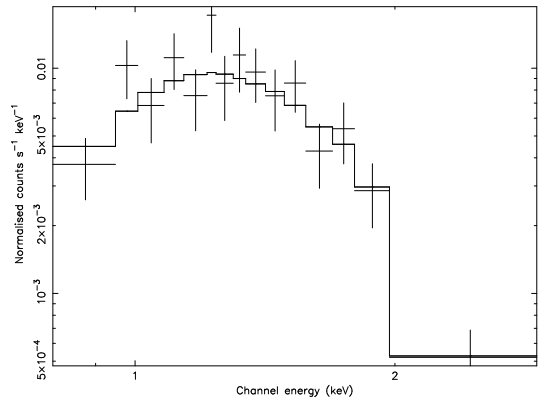}
  
  \caption{The 0.5--10 keV X--ray spectrum of 1M~1716--315 (source 7) in quiescence
  obtained with {\it Chandra}'s ACIS--S on MJD~53567.604 (UTC; July 16, 2005). The solid
  line represents the best--fit absorbed NSA model where N$_H$ is a free--floating fit
  parameter.}

\label{fig:spec}
\end{figure}

The 0.3--10 keV X--ray spectrum was fitted with \textsc {xspec 11.3.1} (\citealt{ar1996}). The fit
function consists of either an absorbed neutron star atmosphere (NSA; \citealt{1991MNRAS.253..193P};
\citealt{1996A&A...315..141Z}), black body or power law model. Each of these absorbed models provides
an acceptable fit to the X--ray spectrum in terms of reduced $\chi^2$. However, the best--fit power law
index was very high with 6.4$^{+1.1}_{-0.8}$ and the interstellar extinction was
1.5$\pm$0.3$\times10^{22}$ cm$^{-2}$ in this case, higher than the best--fit value found in outburst
(errors here and below are at the 68 per cent confidence level). For these reasons we do not consider
the single absorbed power law model any further. We provide the best--fit parameters for the black body
and the NSA model in Table~\ref{tab:fit} (see also Figure~\ref{fig:spec}). In order to determine the
upper limit on the contribution of a powerlaw to the 0.5--10 keV X--ray flux, we determine the
best--fit using a model without a power--law. Next, we fix all the parameters related to this model and
add a power--law to the fit function. The power--law index is fixed to the value 2. Hence, the only
free parameter is the power--law normalisation. After fitting we determine the 90 per cent confidence
error. The outcome of the error in the positive scan direction is added to the value of the power--law
normalisation. The normalisation of the other model components is set to zero as well as the parameter
for the interstellar extinction. Subsequently, the flux in only the power--law component is derived.
Finally, this flux is expressed as a fraction of the flux in the other model components. We give this
95 per cent confidence upper limit in between brackets in Table~\ref{tab:fit}. 

For sources 6, 9  and 10 we only detected 24, 14 and 13 photons in the energy range 0.3--10 keV in the
3\arcsec~source area, respectively. Out of these there are approximately 7$\pm$3 background photons. We
have tried to model the X--ray spectra in \textsc {xspec 11.3.1} making use of the C--statistics,
however, the model parameters are unconstrained. In order to investigate the spectral properties of
these sources, keeping in mind that the background photons generally have a hard spectrum, we determine
the number of photons in the 0.3--1.5 keV and 1.5--10 keV energy band and the mean energy in the
0.3--1.5 keV and 1.5--10 keV bands. For sources 6, 9, and 10 we find 7/17, 2/12 and 2/11 photons in the
0.3--1.5 keV/1.5--10 keV band, respectively. The mean 0.3--1.5 keV/1.5--10 keV energy and the variance
therein for source 6, 9 and 10 is 1.2$\pm$0.3 keV/3.3$\pm$2.2 keV, 1.2$\pm$0.1 keV/4.6$\pm$2.7 keV and
1.0$\pm$0.2 keV/4.4$\pm$2.4 keV, respectively. One can not exclude sources 6, 9 and 10 on the basis of
their hard spectra alone as possible quiescent counterparts of the neutron star SXT 1M~1716-315 since
some neutron stars have been found to have a hard spectrum at low luminosities in quiescence (see
e.g.~\citealt{2004MNRAS.354..666J}). However, only source 7 has a spectrum that is consistent with a
soft spectrum as often found for a quiescent neutron star. Whereas coronal activity from a star might
also yield a soft spectrum (for a review see \citealt{2004A&ARv..12...71G}), we find no evidence in the
optical nor near--infrared for the presence of such a star. From figure 2 in
\citet{2004A&ARv..12...71G}, the observational $I$--band limit and the observed X--ray flux it can be
derived that in order to explain the X--ray emission as coronal X--ray emission of a star, that star
would need to be several orders of magnitude brighter than has been found before in order to explain the
observed X--ray flux and the optical non--detection. This makes the identification of the X--ray source
as stellar coronal X--ray emission unlikely. Instead, we conclude that based on the positional
coincidence, the spectral properties together with the optical and near--infrared upper limits, it is
very likely that source 7 is the quiescent neutron star counterpart to 1M~1716--315.

\begin{table*}

\caption{Best fit parameters of the quiescent spectrum of 1M~1716--315. NSA stands for neutron star atmosphere and BB refers to
blackbody.  All quoted errors are at the 68 per cent confidence level. The value in between brackets in the unabsorbed flux column
denotes the 95 per cent upper limit to the fractional contribution of a power law (the power law index was fixed at 2; see text). }

\label{tab:fit}
\begin{tabular}{lccccccl}
\hline
Model & N$_H$$\times10^{21}$& BB radius & NSA norm. & Temp. & Unabs.~0.5--10 keV flux & Abs.~0.5--10 keV flux &${\rm \chi^2_{red}}$ \\ 
      & cm$^{-2}$ & ${\rm (\frac{d}{10\,kpc})^2}$ km & d (kpc) & BB (keV)/NSA & (erg cm$^{-2}$ s$^{-1}$) & (erg cm$^{-2}$ s$^{-1}$) &d.o.f.\\
      & &  &  & log K/keV$^\infty$ &  &  &  \\
\hline
\hline
BB  & 2.1$^a$ & 0.4$\pm 0.1$ & -- & 0.33$\pm$0.02& $(5.0^{+0.3}_{-0.8})\times 10^{-14}$ (19\%)& $(3.3^{+0.2}_{-0.4})\times 10^{-14}$ & 1.16/13 \\
NSA & 2.1$^a$ & -- & 53$\pm$11& 6.42$\pm$0.04/0.17$\pm$0.02 & $(5.4^{+2.2}_{-2.6})\times 10^{-14}$ (18\%)& $(3.7^{+1.2}_{-1.6})\times 10^{-14}$& 1.39/13 \\
\hline
BB & 7$\pm$2 &  5$^{+9}_{-5}$& --  & 0.24$\pm 0.03$ & $(1.3^{+0.1}_{-0.6})\times 10^{-13}$ (16\%) & $(3.1^{+0.1}_{-1.4})\times 10^{-14}$ & 0.64/12\\
NSA & 9$\pm$2 & --  & 3.9$^{+b}_{-2.3}$ & 6.07$\pm$0.08/0.078$\pm$0.014  & $(2.4^b)\times 10^{-13}$ (8\%)& $(3.2^b)\times 10^{-14}$  & 0.63/12 \\
\end{tabular}

{\footnotesize$^a$ Parameter fixed.}\\
{\footnotesize$^b$ Error range is unconstrained.}\\
\end{table*}

With the new accurate {\it Chandra} X--ray position from our ACIS--S observation in hand, we
re--analysed the HRC--I observation of Wachter et al.~(2005). Only one photon is detected in the
0\farcs67 90 per cent confidence error circle centred on the best--fit source position (taking
into account the nominal 0\farcs6 90 per cent error for {\it Chandra} observations since we
could not correct the bore sight for the {\it Chandra}--HRC observation). Using
\citet{1986ApJ...303..336G} this gives a 95 per cent confidence upper limit on the source count
rate of 4$\times 10^{-3}$ counts s$^{-1}$. Assuming the same spectral energy distribution as
found from our {\it Chandra} ACIS--S observation and fixing N$_H$ to 2.1$\times 10^{21}$
cm$^{-2}$, the 95 per cent confidence upper limit on the unabsorbed 0.5--10 keV source flux is
5.6$\times 10^{-14}$ erg cm$^{-2}$ s$^{-1}$, slightly higher than that derived by Wachter et
al.~(2005; in this we used \textsl {w3pimms}\footnote{Available at
http://heasarc.gsfc.nasa.gov/Tools/w3pimms.html}).  The upper limit on the flux derived from the
{\it Chandra} HRC--I observation is consistent with the flux measured in our ACIS--S
observation. 

\section{Discussion}

We have observed the field of the neutron star SXT 1M~1716--315 in X--rays with the {\it Chandra}
satellite for $\sim20$ ksec. We detect only one source in the HEAO--I error region. The spectrum of
the source is soft with a 95 per cent upper limit to a power law contribution of $<$19 per cent
making it unlikely that this source is a background AGN or an accreting white dwarf. The X--ray
luminosity together with our stringent optical $I$--band and near--infrared $K_s$--band limits on
the absolute magnitude rule out that the soft X--ray spectrum is due to coronal activity of a late
type star. Three other sources were found near the 90 per cent confidence HEAO--I error region.
Even though the detected number of counts for these sources is too low to fit a spectrum nearly all
photons of all three sources are detected above 1.5 keV. Together with the fact that their
positions do not agree with the HEAO--I error region, this makes them less likely candidates for
the quiescent neutron star SXT. 

We conclude that we have detected 1M~1716--315 in quiescence at a 0.5--10 keV source luminosity of
$1.6\times10^{32}$--$1.3\times10^{33}$ erg s$^{-1}$. Fixing the value for the interstellar N$_H$ to
the value given in \citet{1997ApJS..109..177C} [(2.1$\pm$0.6)$\times$$10^{21}$ cm$^{-2}$] the
resultant blackbody and neutron star atmosphere temperature is rather high for a neutron star SXT
in quiescence, especially when taking into account that the source has been in quiescence for more
than $\approx$16 years (see below). We note that the N$_H$ value given in table 12 of
\citet{1997ApJS..109..177C} depends on the spectral model that was used, both an absorbed blackbody
plus thermal bremsstrahlung as well as Comptonisation models give a N$_H$ value of
4--5$\times$10$^{21}$ cm$^{-2}$ (see tables 7 and 10 in \citealt{1997ApJS..109..177C}). Indeed,
when left as a free parameter in our X--ray spectral fits the N$_H$ is somewhat higher, although
still consistent within 3~$\sigma$ with the value of (2.1$\pm$0.6)$\times$$10^{21}$ cm$^{-2}$. 

The $I$--band observations do not reveal an optical counterpart in the 99 per cent confidence 0\farcs36
error circle down to a magnitude limit of $I>23.5$. Using this limit and converting the N$_H$ observed
in outburst giving A$_I\sim 0.7$ (\citealt{scfida1998}), taking d=5.1--6.9 kpc
(\citealt{2004MNRAS.354..355J}) we derive that the absolute $I$--band magnitude M$_I\geq8.6-9.3$.
Equivalent calculations for the $K_s$ band measurements result in  $K_s$$\geq$5.1--5.8. Hence, like
1H~1905+000 (\citealt{2006MNRAS.tmp..395J}) 1M~1716--315 is a candidate UCXB, although the secondary
could also be an M dwarf. However, additional (circumstantial) evidence for an ultra--compact nature
comes from the detected long ($\approx$10 minutes) type I burst duration
(\citealt{1976ApJ...208L.115M}). Such long bursts can occur in sources accreting He at a low rate in
ultra--compact X--ray binaries (see discussion in \citealt{2007inpress}) or sources accreting H--rich
material at even lower rates (\citealt{2007ApJ...654.1022P}). Since the luminosity in outburst in the
case of 1M~1716--315 was found to be above 10$^{36}$ erg s$^{-1}$ the latter mode is less likely. The
quiescent X--ray luminosity of 1M~1716--315 is higher than that of other (likely) UCXBs in quiescence
(XTE~J1751--305, XTE~J0929--314, XTE~J1807--294, and 1H~1905+000 ; \citealt{2005ApJ...619..492W};
\citealt{2005A&A...434L...9C}; \citealt{2006MNRAS.tmp..395J}). A plausible explanation for the higher
quiescent luminosity is that the time--averaged mass accretion rate is higher in 1M~1716--315 than in
the other UCXBs observed so far, possibly due to a shorter orbital period
(\citealt{2003ApJ...598.1217D}).

\begin{figure}
  \includegraphics[angle=0,width=8cm,clip]{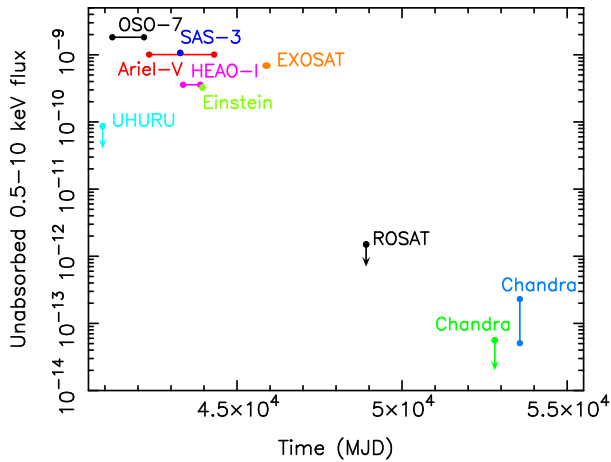}
  
  \caption{The unabsorbed 0.5--10 keV X--ray flux history of the source 1M~1716--315. The
plotted ranges for each satellite indicate the time span over which the source was detected on
multiple occasions by that satellite. In order to convert fluxes given in the literature in
other X--ray bands to the 0.5--10 keV band we assume the spectrum to be well--represented by an
absorbed power law with index 2 (except for the ROSAT and {\it Chandra} upper limit which were
derived assuming a 0.3 keV black body). The absorption was taken to be 2.1$\times 10^{21}$
cm$^{-2}$ as found from spectra obtained in outburst by Einstein
(\citealt{1997ApJS..109..177C}). Arrows on the data points indicate upper limits to the X--ray
flux.}

\label{fig:lc}
\end{figure}

We have plotted the long--term lightcurve of 1M~1716-315 in Figure~\ref{fig:lc}.\footnote{Data for
Figure~\ref{fig:lc} was taken or derived from \citet{1978ApJS...38..357F} (UHURU);
\citet{1975IAUC.2765Q...1M}, \citet{1976ApJ...208L.115M} and \citet{1977ApJ...218..801M} (OSO--7),
\citet{1978Natur.272..701J} (SAS--3), \citet{1980AJ.....85.1062R} (HEAO--I), \citet{1997ApJS..109..177C}
(Einstein), \citet{1988MNRAS.232..551W} and \citet{1995A&AS..109....9G} (EXOSAT), 
\citet{1999A&A...349..389V} (ROSAT), and this work}. The most striking feature is the longevity of the
outburst. The source has been active for nearly 13 years before it returned to quiescence
(Fig.~\ref{fig:lc}). There is a time span of 8.2 years between the last detection by EXOSAT and the first
non--detection by ROSAT. The source is likely to have returned to quiescence during this interval. There
is 16.9 years between the mid--point of this interval and the time of the {\it Chandra} observation that
detected the source. Hence, the source has been in quiescence for 16.9$\pm$4.1 years. There is a growing
number of quasi--persistent sources (the other sources being 2S~1711--339\footnote{This source has had a
$\approxgt$3 year--long outburst between MJD~50,739 and 51,751. On MJD~52,345 it was not detected with
{\it Chandra} (see also \citealt{2003ApJ...596.1220W}).}, 4U~2129+47, KS~1731--260, MXB~1659--290,
X~1732--304, and 1H~1905+000, \citealt{2004cxo..prop.1624W}; \citealt{2004ATel..233....1T};
\citealt{2006MNRAS.tmp..479C}; \citealt{2006astro.ph..5490C}; \citealt{2006MNRAS.tmp..395J}). Of these,
only 1H~1905+000 has so far not been detected in quiescence.  At present there are three other
quasi--persistent sources that are still in outburst, EXO~0748--676, GS~1826--238, and XTE~J1759--220. In
addition, HETE~J1900.1--2455 has been in outburst for more than a year now. All these quasi--persistent
sources have confirmed neutron star accretors, except XTE~J1759--220 for which type~I bursts or
pulsations have not been reported so far.

\section*{Acknowledgments}  

\noindent PGJ and SW acknowledges support from NASA grant GO4-5033X. PGJ and CGB acknowledge
support from the Netherlands Organisation for Scientific Research. We also acknowledge Frank
Verbunt for useful discussions and the referee for useful comments that helped improve the
manuscript.


\begin{thebibliography}{42}
\expandafter\ifx\csname natexlab\endcsname\relax\def\natexlab#1{#1}\fi

\bibitem[{{Arnaud}(1996)}]{ar1996}
{Arnaud}, K.~A., 1996, in ASP Conf. Ser. 101: Astronomical Data Analysis
  Software and Systems V, vol.~5, p.~17

\bibitem[{{Belczynski} \& {Taam}(2004)}]{2004ApJ...603..690B}
{Belczynski}, K., {Taam}, R.~E., 2004, \apj, 603, 690

\bibitem[{{Brown} et~al.(1998){Brown}, {Bildsten}, \&
  {Rutledge}}]{1998ApJ...504L..95B}
{Brown}, E.~F., {Bildsten}, L., {Rutledge}, R.~E., 1998, \apjl, 504, L95

\bibitem[{{Cackett} et~al.(2006{\natexlab{a}}){Cackett}, {Wijnands}, {Linares},
  {Miller}, {Homan}, \& {Lewin}}]{2006astro.ph..5490C}
{Cackett}, E.~M., {Wijnands}, R., {Linares}, M., {Miller}, J.~M., {Homan}, J.,
  {Lewin}, W., 2006{\natexlab{a}}, ArXiv Astrophysics e-prints

\bibitem[{{Cackett} et~al.(2006{\natexlab{b}}){Cackett}, {Wijnands}, {Linares},
  {Miller}, {Homan}, \& {Lewin}}]{2006MNRAS.372..479C}
{Cackett}, E.~M., {Wijnands}, R., {Linares}, M., {Miller}, J.~M., {Homan}, J.,
  {Lewin}, W.~H.~G., 2006{\natexlab{b}}, \mnras, 372, 479

\bibitem[{{Cackett} et~al.(2006{\natexlab{c}})}]{2006MNRAS.tmp..479C}
{Cackett}, E.~M., et~al., 2006{\natexlab{c}}, \mnras, 479

\bibitem[{{Campana} et~al.(2005){Campana}, {Ferrari}, {Stella}, \&
  {Israel}}]{2005A&A...434L...9C}
{Campana}, S., {Ferrari}, N., {Stella}, L., {Israel}, G.~L., 2005, \aap, 434,
  L9

\bibitem[{{Chen} et~al.(1997){Chen}, {Shrader}, \&
  {Livio}}]{1997ApJ...491..312C}
{Chen}, W., {Shrader}, C.~R., {Livio}, M., 1997, \apj, 491, 312

\bibitem[{{Christian} \& {Swank}(1997)}]{1997ApJS..109..177C}
{Christian}, D.~J., {Swank}, J.~H., 1997, \apjs, 109, 177

\bibitem[{{Colpi} et~al.(2001){Colpi}, {Geppert}, {Page}, \&
  {Possenti}}]{2001ApJ...548L.175C}
{Colpi}, M., {Geppert}, U., {Page}, D., {Possenti}, A., 2001, \apjl, 548, L175

\bibitem[{{Deloye} \& {Bildsten}(2003)}]{2003ApJ...598.1217D}
{Deloye}, C.~J., {Bildsten}, L., 2003, \apj, 598, 1217

\bibitem[{{Forman} et~al.(1978){Forman}, {Jones}, {Cominsky}, {Julien},
  {Murray}, {Peters}, {Tananbaum}, \& {Giacconi}}]{1978ApJS...38..357F}
{Forman}, W., {Jones}, C., {Cominsky}, L., {Julien}, P., {Murray}, S.,
  {Peters}, G., {Tananbaum}, H., {Giacconi}, R., 1978, \apjs, 38, 357

\bibitem[{{Gehrels}(1986)}]{1986ApJ...303..336G}
{Gehrels}, N., 1986, \apj, 303, 336

\bibitem[{{Gottwald} et~al.(1995){Gottwald}, {Parmar}, {Reynolds}, {White},
  {Peacock}, \& {Taylor}}]{1995A&AS..109....9G}
{Gottwald}, M., {Parmar}, A.~N., {Reynolds}, A.~P., {White}, N.~E., {Peacock},
  A., {Taylor}, B.~G., 1995, \aaps, 109, 9

\bibitem[{{G{\"u}del}(2004)}]{2004A&ARv..12...71G}
{G{\"u}del}, M., 2004, \araa, 12, 71

\bibitem[{{in't Zand} et~al.(2007){in't Zand}, {Jonker}, \&
  {Markwardt}}]{2007inpress}
{in't Zand}, J.~J.~M., {Jonker}, P.~G., {Markwardt}, C.~B., 2007, \aa

\bibitem[{{Jernigan} et~al.(1978){Jernigan}, {Bradt}, {Doxsey}, {Dower},
  {McClintock}, \& {Apparao}}]{1978Natur.272..701J}
{Jernigan}, J.~G., {Bradt}, H.~V., {Doxsey}, R.~E., {Dower}, R.~G.,
  {McClintock}, J.~E., {Apparao}, K.~M.~V., 1978, \nat, 272, 701

\bibitem[{{Jonker} \& {Nelemans}(2004)}]{2004MNRAS.354..355J}
{Jonker}, P.~G., {Nelemans}, G., 2004, \mnras, 354, 355

\bibitem[{{Jonker} et~al.(2004){Jonker}, {Galloway}, {McClintock}, {Buxton},
  {Garcia}, \& {Murray}}]{2004MNRAS.354..666J}
{Jonker}, P.~G., {Galloway}, D.~K., {McClintock}, J.~E., {Buxton}, M.,
  {Garcia}, M., {Murray}, S., 2004, \mnras, 354, 666

\bibitem[{{Jonker} et~al.(2006){Jonker}, {Bassa}, {Nelemans}, {Juett}, {Brown},
  \& {Chakrabarty}}]{2006MNRAS.tmp..395J}
{Jonker}, P.~G., {Bassa}, C.~G., {Nelemans}, G., {Juett}, A.~M., {Brown},
  E.~F., {Chakrabarty}, D., 2006, \mnras, 395

\bibitem[{{Makishima} et~al.(1981)}]{1981ApJ...244L..79M}
{Makishima}, K., et~al., 1981, \apjl, 244, L79

\bibitem[{{Markert} et~al.(1975){Markert}, {Bradt}, {Clark}, {Lewin}, {Li},
  {Schnopper}, {Sprott}, \& {Wargo}}]{1975IAUC.2765Q...1M}
{Markert}, T.~H., {Bradt}, H.~V., {Clark}, G.~W., {Lewin}, W.~H.~G., {Li},
  F.~K., {Schnopper}, H.~W., {Sprott}, G.~F., {Wargo}, G.~F., 1975, \iaucirc,
  2765, 1

\bibitem[{{Markert} et~al.(1976){Markert}, {Backman}, \&
  {McClintock}}]{1976ApJ...208L.115M}
{Markert}, T.~H., {Backman}, D.~E., {McClintock}, J.~E., 1976, \apjl, 208, L115

\bibitem[{{Markert} et~al.(1977){Markert}, {Canizares}, {Clark}, {Hearn}, {Li},
  {Sprott}, \& {Winkler}}]{1977ApJ...218..801M}
{Markert}, T.~H., {Canizares}, C.~R., {Clark}, G.~W., {Hearn}, D.~R., {Li},
  F.~K., {Sprott}, G.~F., {Winkler}, P.~F., 1977, \apj, 218, 801

\bibitem[{{Pavlov} et~al.(1991){Pavlov}, {Shibanov}, \&
  {Zavlin}}]{1991MNRAS.253..193P}
{Pavlov}, G.~G., {Shibanov}, I.~A., {Zavlin}, V.~E., 1991, \mnras, 253, 193

\bibitem[{{Peng} et~al.(2007){Peng}, {Brown}, \&
  {Truran}}]{2007ApJ...654.1022P}
{Peng}, F., {Brown}, E.~F., {Truran}, J.~W., 2007, \apj, 654, 1022

\bibitem[{{Reid} et~al.(1980){Reid}, {Johnston}, {Bradt}, {Doxsey},
  {Griffiths}, \& {Schwartz}}]{1980AJ.....85.1062R}
{Reid}, C.~A., {Johnston}, M.~D., {Bradt}, H.~V., {Doxsey}, R.~E., {Griffiths},
  R.~E., {Schwartz}, D.~A., 1980, \aj, 85, 1062

\bibitem[{{Rutledge} et~al.(2002){Rutledge}, {Bildsten}, {Brown}, {Pavlov},
  {Zavlin}, \& {Ushomirsky}}]{2002ApJ...580..413R}
{Rutledge}, R.~E., {Bildsten}, L., {Brown}, E.~F., {Pavlov}, G.~G., {Zavlin},
  V.~E., {Ushomirsky}, G., 2002, \apj, 580, 413

\bibitem[{{Schlegel} et~al.(1998){Schlegel}, {Finkbeiner}, \&
  {Davis}}]{scfida1998}
{Schlegel}, D.~J., {Finkbeiner}, D.~P., {Davis}, M., 1998, \apj, 500, 525

\bibitem[{{Tawara} et~al.(1984{\natexlab{a}}){Tawara}, {Hirano}, {Kii},
  {Matsuoka}, \& {Murakami}}]{1984PASJ...36..861T}
{Tawara}, Y., {Hirano}, T., {Kii}, T., {Matsuoka}, M., {Murakami}, T.,
  1984{\natexlab{a}}, \pasj, 36, 861

\bibitem[{{Tawara} et~al.(1984{\natexlab{b}})}]{1984ApJ...276L..41T}
{Tawara}, Y., et~al., 1984{\natexlab{b}}, \apjl, 276, L41

\bibitem[{{Torres} et~al.(2004){Torres}, {McClintock}, {Garcia}, \&
  {Murray}}]{2004ATel..233....1T}
{Torres}, M.~A.~P., {McClintock}, J.~E., {Garcia}, M.~R., {Murray}, S.~S.,
  2004, The Astronomer's Telegram, 233, 1

\bibitem[{{Verbunt} \& {van den Heuvel}(1995)}]{verbunt1995}
{Verbunt}, F., {van den Heuvel}, E., 1995, Formation andevolution of neutron
  stars and black holes in binaries, eds Lewin, van Paradijs, van den Heuvel,
  ISBN 052141684, Cambridge University Press, 1995.

\bibitem[{{Voges} et~al.(1999)}]{1999A&A...349..389V}
{Voges}, W., et~al., 1999, \aap, 349, 389

\bibitem[{{Wachter} et~al.(2005){Wachter}, {Wellhouse}, \&
  {Bandyopadhyay}}]{2005AIPC..797..639W}
{Wachter}, S., {Wellhouse}, J.~W., {Bandyopadhyay}, R.~M., 2005, in {Burderi},
  L., {Antonelli}, L.~A., {D'Antona}, F., {di Salvo}, T., {Israel}, G.~L.,
  {Piersanti}, L., {Tornamb{\`e}}, A., {Straniero}, O., eds., AIP Conf. Proc.
  797: Interacting Binaries: Accretion, Evolution, and Outcomes, p. 639

\bibitem[{{Warwick} et~al.(1988){Warwick}, {Norton}, {Turner}, {Watson}, \&
  {Willingale}}]{1988MNRAS.232..551W}
{Warwick}, R.~S., {Norton}, A.~J., {Turner}, M.~J.~L., {Watson}, M.~G.,
  {Willingale}, R., 1988, \mnras, 232, 551

\bibitem[{{Wijnands}(2004)}]{2004cxo..prop.1624W}
{Wijnands}, R., 2004, in Crust cooling curves of accretion-heated neutron
  stars, p. 1624

\bibitem[{{Wijnands} et~al.(2001){Wijnands}, {Miller}, {Markwardt}, {Lewin}, \&
  {van der Klis}}]{2001ApJ...560L.159W}
{Wijnands}, R., {Miller}, J.~M., {Markwardt}, C., {Lewin}, W.~H.~G., {van der
  Klis}, M., 2001, \apjl, 560, L159

\bibitem[{{Wijnands} et~al.(2005){Wijnands}, {Homan}, {Heinke}, {Miller}, \&
  {Lewin}}]{2005ApJ...619..492W}
{Wijnands}, R., {Homan}, J., {Heinke}, C.~O., {Miller}, J.~M., {Lewin},
  W.~H.~G., 2005, \apj, 619, 492

\bibitem[{{Wilson} et~al.(2003){Wilson}, {Patel}, {Kouveliotou}, {Jonker}, {van
  der Klis}, {Lewin}, {Belloni}, \& {M{\'e}ndez}}]{2003ApJ...596.1220W}
{Wilson}, C.~A., {Patel}, S.~K., {Kouveliotou}, C., {Jonker}, P.~G., {van der
  Klis}, M., {Lewin}, W.~H.~G., {Belloni}, T., {M{\'e}ndez}, M., 2003, \apj,
  596, 1220

\bibitem[{{Zacharias} et~al.(2004){Zacharias}, {Urban}, {Zacharias}, {Wycoff},
  {Hall}, {Monet}, \& {Rafferty}}]{2004AJ....127.3043Z}
{Zacharias}, N., {Urban}, S.~E., {Zacharias}, M.~I., {Wycoff}, G.~L., {Hall},
  D.~M., {Monet}, D.~G., {Rafferty}, T.~J., 2004, \aj, 127, 3043

\bibitem[{{Zavlin} et~al.(1996){Zavlin}, {Pavlov}, \&
  {Shibanov}}]{1996A&A...315..141Z}
{Zavlin}, V.~E., {Pavlov}, G.~G., {Shibanov}, Y.~A., 1996, \aap, 315, 141

\end{thebibliography}
\end{document}